\documentclass[aps,prb,twocolumn,superscriptaddress,amsmath]{revtex4-2}
\usepackage[colorlinks=true, linkcolor=blue, filecolor=blue, urlcolor=blue, citecolor=blue]{hyperref}
\usepackage{amsfonts}
\usepackage{bbm}
\usepackage{graphicx}
\usepackage{dcolumn}
\usepackage{bm}
\usepackage{soul}

\begin{document}

\title{A Design of Josephson Diode Based on Magnetic Impurity}

\author{Yu-Fei Sun}
\email[]{These authors contributed equally to this work.}
\affiliation{International Center for Quantum Materials, School of Physics, Peking University, Beijing 100871, China}
\affiliation{Hefei National Laboratory, Hefei 230088, China}

\author{Yue Mao}
\email[]{These authors contributed equally to this work.}
\affiliation{International Center for Quantum Materials, School of Physics, Peking University, Beijing 100871, China}

\author{Qing-Feng Sun}
\email[]{sunqf@pku.edu.cn}
\affiliation{International Center for Quantum Materials, School of Physics, Peking University, Beijing 100871, China}
\affiliation{Hefei National Laboratory, Hefei 230088, China}
\affiliation{Collaborative Innovation Center of Quantum Matter, Beijing 100871, China}

\begin{abstract}
We theoretically propose a mechanism to realize the superconducting diode effect (SDE):
The current can generate a magnetic field, affecting the magnetic moment of magnetic impurity.
When the connection region of the Josephson junction is coupled with the magnetic impurity, the supercurrents in positive and negative directions have different influences on the magnetic moment.
This results in a phenomenon that the critical supercurrents in these opposite directions are unequal, which is called SDE.
We model the Josephson connection region by a quantum dot.
Then the critical supercurrents are investigated by the non-equilibrium Green's function method, and we carry out a detailed symmetry analysis on the supercurrent relations.
The calculation results confirm that the SDE does exist in this system.
Besides, the SDE is significant in a wide parameter space and can be effectively adjusted in various ways.
Our design only demands a magnetic impurity and conventional superconductors.
The unconventional finite-momentum Cooper pair and spin-orbit coupling are not required, and there is also no need for the existence of chirality or an external magnetic field.
Our work provides a universal device structure for the development of superconducting electronics.

\end{abstract}
\maketitle

\section{\label{sec1}Introduction}

As one of the most commonly used components in modern electronic technologies, semiconductor diodes play a significant role in the fields of computation and electronic devices.
This is due to their characteristics of having high resistance in one direction and low resistance in the opposite direction.
However, heat generation and energy dissipation are inevitable in
the traditional semiconductor diode, due to the presence of non-zero resistance.
To fabricate faster and more energy-efficient devices, scientists have proposed the research goal of achieving the superconducting diode effect (SDE) \cite{p1,p2,p3,p4,p5,p6,p7,p8,p9,p10,p11,p12,p13,p14,p15,p16,p17,p18,p19,p20,p21,p22,p23,p24,p25,p26,p27,p28,p29,p30,p31,p32},
which have unidirectional nondissipative transport and zero resistance in a single direction. In addition, SDEs also lead the development of direction-selective quantum sensors and superconducting quantum computing qubits \cite{p9,p19}.

Currently, the realization of the bulk SDE usually relies on unconventional superconducting properties.
By utilizing the so-called magnetochiral anisotropy effect \cite{p33,p34,p35,p36,p37}, the SDE can be achieved by applying a magnetic field to superconductors with spin-orbit coupling \cite{p1,p2,p11,p12,p13,p14,p15,p16}.
Theoretically, the combination of spin-orbit coupling and magnetic field can cause a finite Cooper pair momentum, which is regarded as the origin of the bulk SDE \cite{p4,p5,p6,p11,p12,p13,p14,p15,p16,p17}.
In addition to bulk superconductors, SDEs can also be achieved in Josephson junctions \cite{p5,p6,p7,p8,p9,p10,p18,p19,p20,p21,p22,p23,p24,p25,p26,p27,p28,p29,p30,p31,p32}, which are also called Josephson diodes.
Theoretical works have calculated anomalous current-superconducting phase relationships when the spin-orbit coupling and magnetism both exist in the connection region of the Josephson junction, from which the phenomenon of unequal critical currents in opposite directions can be seen.
Here, the connection region can be quantum dots \cite{p20,p21}, nanowires \cite{p22}, metals \cite{p23}, topological insulators \cite{p24}, and two-dimensional electron gases \cite{p25}.
Furthermore, the Josephson diode effect can also be realized by constructing asymmetric SQUIDs \cite{p28,p29}, using chiral material as the connection region \cite{p30}, or utilizing the valley polarization \cite{p31}. Besides, the Josephson diode effects have also been successfully implemented in experiments \cite{p5,p6,p7,p8,p9,p10}.

At present, the main scheme for realizing the SDE almost depends on the finite Cooper pair momentum, strong spin-orbit coupling, magnetic field, chirality, etc.
However, these special requirements will limit the practical application range of the superconducting diodes.
Therefore, researchers are still looking for more applicable superconducting diodes, e.g., the field-free superconducting diodes \cite{p3,p4,p7,p9}. Very recently, it was found that the SDE can even be induced by a magnetic atom:
Trahms et al. experimentally observed both direction-dependent critical current and retrapping current in a Josephson junction connected by a magnetic atom \cite{p10}.
They focus on the nonreciprocal retrapping current, which was understood by the asymmetric quasiparticle tunneling spectra of Yu-Shiba-Rusinov states \cite{p10,p32}.
But the critical current, corresponding to the current-phase relation, is rather vague in the mechanism. How do magnetic atoms cause Josephson diodes?
The principle behind it should be able to universally guide the future studies on Josephson diode.

In this paper, we theoretically propose a concise and general Josephson diode device based on magnetic atoms: a Josephson junction with its \emph{connection region coupled to a magnetic impurity}.
When a supercurrent flows through the Josephson junction, a magnetic field is generated, which in turn affects the magnetic moment of magnetic impurity.
Because the currents in opposite directions have different effects
on the magnetic moment, the critical currents in these opposite directions
are unequal, and the SDE can be achieved.
Using the non-equilibrium Green's function method,
we calculate the Josephson current of the system and confirm the existence of SDE.
We systematically study the influence of system's parameters on the superconducting diode efficiency, such as magnetic moment,
coupling strength, and energy level.
We also carry out a detailed symmetry analysis on the current relations.
The results show that the SDE does exist and is significant in a wide range of parameters, where the efficiency can reach 0.5.
The SDE can be regulated by controlling gate voltage, coupling strength, etc.
{Although our proposal is more for fundamental physical content than mature industrial applications,
the idea and mechanism is novel with} the following advantages:
Firstly, the finite-momentum Cooper pairs in unconventional superconductors
are not required, and the SDE can just be realized by conventional $s$-wave superconductors;
Secondly, the spin-orbit coupling, as a key factor of numerous researches on SDE, is not required in our design;
Thirdly, there is no need for an external magnetic field or chirality.
Therefore, our proposal has high practicability and gives a new perspective on the construction of superconducting circuits, {especially for the urgent nanoscale three-terminal superconducting devices that can directly reduce thermal dissipation in logic circuits \cite{BookSD}.}

The rest of this paper is organized as follows:
In Sec. \ref{sec2}, we present the model Hamiltonian of our Josephson diode device and show the method to calculate the Josephson current.
In Sec. \ref{sec3}, we confirm the SDE by calculation and explain its underlying mechanism.
In Sec. \ref{sec4}, we systematically study the regulation of various parameters on the SDE.
Finally, discussion and conclusion are given in Sec. \ref{sec5}.

\section{\label{sec2}Model and formula}

\begin{figure}[!htb]
\centerline{\includegraphics[width=\columnwidth]{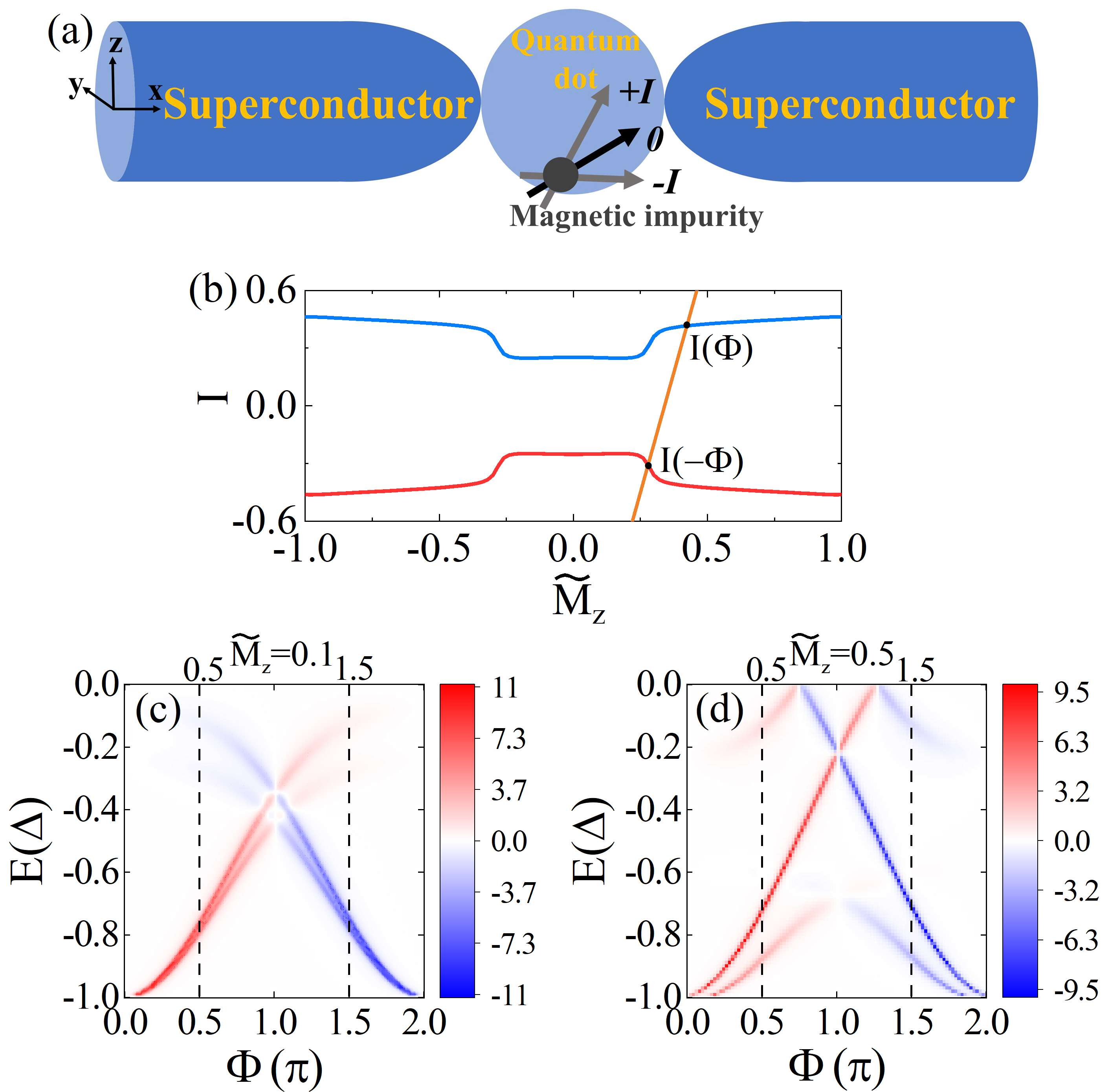}}
\caption{(a) The schematic diagram of the Josephson diode model, which consists of two superconductor leads and a connection region (modelled by a QD) coupling to the magnetic impurity.
The magnetic fields generated by the positive and negative supercurrents $I$ and $-I$ have different effects on the magnetic moment of the magnetic impurity, resulting in the SDE.
(b) The blue and red curves are the plots of the supercurrent $I$-effective magnetic moment $\widetilde{M_z}$ relations under the superconducting phase difference $\phi$ and $-\phi$, respectively.
The yellow line represents $I=\left(\widetilde{M_z}-M_z\right)/\alpha$ from the relation $\widetilde{M_z}=M_z+\alpha I$.
The intersections of the yellow line with the blue curve and the red curve are the self-consistently calculated currents under the phase difference $\phi$ and $-\phi$, respectively.
The combined effect of $M_z$ and $\alpha$ leads to the result that $I\left(-\phi\right)\neq-I(\phi)$.
(c, d) The energy distribution diagram of the integrand $i(\epsilon)$ of the supercurrent with $\widetilde{M_z}=0.1,0.5$, respectively. The dashed lines mark the source of the currents corresponding to (b) with $\phi$ and $-\phi$. Parameters: (b-d) $\varepsilon_D=\varepsilon_M=0$, $t_c=1$. (b) $\phi=0.5\pi$.}\label{Fig1}
\end{figure}

The proposed Josephson diode device consists of two superconductor leads with their connection region coupled to a magnetic impurity, as shown in Fig. {\ref{Fig1}}(a). For simplicity, we use a single-level quantum dot (QD) to represent the connection region between the two superconductor leads. The general Hamiltonian can be written as \cite{p38,p39,p42}
\begin{eqnarray}
H=\sum_{\beta=L,R} H_\beta+H_{mid}+H_t \label{E1},
\end{eqnarray}
where $H_\beta(\beta=L, R)$ is the Hamiltonian of the left or right superconductor lead, $H_{mid}$ describes the central QD and the magnetic impurity, $H_t$ describes the coupling between QD and superconductors, as well as the coupling between QD and the magnetic impurity.
$H_\beta$, $H_{mid}$, and $H_t$ are respectively expressed as
\begin{eqnarray}
H_\beta &=& \sum_{k\sigma} \varepsilon_k C_{\beta,k\sigma}^\dagger C_{\beta,k\sigma}\nonumber \\
&+&\sum_{k} \left( \Delta e^{i\phi_\beta} C_{\beta,-k\uparrow}^\dagger C_{\beta,k\downarrow}^\dagger+H.C.\right), \label{E2} \\
H_{mid} &=& \sum_{\sigma}\varepsilon_D d_\sigma^\dagger d_\sigma \nonumber\\
&+&\sum_{\sigma}\ \begin{pmatrix} a_\uparrow^\dagger & a_\downarrow^\dagger \end{pmatrix} \left(\varepsilon_M+\widetilde{\mathbf{M}} \cdot \vec{\sigma}\right)   \begin{pmatrix} a_\uparrow \\ a_\downarrow\\\end{pmatrix}, \label{E3} \\
H_t &=& \left(\sum_{k,\sigma}\ t_s C_{L,k\sigma}^\dagger d_\sigma+t_s C_{R,k\sigma}^\dagger d_\sigma+H.C.\right)\nonumber \\
&+&\sum_{\sigma}\ t_c(d_\sigma^\dagger a_\sigma+H.C.). \label{E4}
\end{eqnarray}
Here $C_{\beta,k\sigma}^\dagger$, $d_\sigma^\dagger$ and $a_\sigma^\dagger$ ($C_{\beta,k\sigma}$, $d_\sigma$ and $a_\sigma$) are the creation (annihilation) operators of electrons in the superconductors, QD, and magnetic impurity, respectively. $\sigma=\uparrow$, $\downarrow$ represents the spin,
$\vec{\sigma}=(\sigma_x,\sigma_y,\sigma_z)$ is the Pauli matrices in spin space,
$\Delta$ is the superconducting gap and $\phi_\beta$ is the superconducting phase of the $\beta$ side.
We set $\phi_L=0$, $\phi_R=\phi$ with a phase difference $\phi$.
$\varepsilon_D$ and $\varepsilon_M$ are the energy levels of QD and magnetic impurity, respectively. $t_s$ and $t_c$ are the coupling strength between QD and superconductors, and the coupling strength between QD and the magnetic impurity, respectively.
$\widetilde{\mathbf{M}}$ represents the Zeeman splitting from the magnetic
moment of magnetic impurity.
Because that all the other terms in the Hamiltonian are independent
of the choice of spin quantization axis,
we can set $\widetilde{\mathbf{M}}=(0,0,\widetilde{M}_z)$ being along the $z$ direction.
When the Josephson current $I$ flows along the $x$ direction,
a magnetic field in the $z$ direction is induced around the magnetic impurity,
which will affect the magnetic moment $\widetilde{M}_z$.
We assume the magnetic moment $\widetilde{M}_z=M_z+\alpha I$,
where $M_z$ is the intrinsic magnetic moment of the magnetic impurity at zero current
and the term $\alpha I$ is the extra magnetic moment from the Josephson supercurrent $I$.
Here the extra magnetic moment is proportional to the current with the coefficient $\alpha$, of which the value will be discussed in Sec. \ref{sec4}.
The current $I$ can be calculated via the Hamiltonian.
Since $I$ in turn influences the Hamiltonian, the value of $I$ should be calculated self-consistently.

The Josephson supercurrent can be calculated from the evolution of the particle number operator of the electrons in the left superconductor \cite{p40}
\begin{align}
    I&=-e\left\langle\frac{d}{dt}\sum_{k\sigma}{C_{L,k\sigma}^\dag C_{L,k\sigma}}\right\rangle\nonumber\\
    &=\frac{2e}{\hbar}\left\{Re\left[\ t_sG_{DL,11}^<\left(t,t\right)\right]+Re\left[\ t_sG_{DL,22}^<\left(t,t\right)\right]\right\},	\label{E5}
\end{align}
where $G_{DL,11}^<\left(t,t\right)$ and $G_{DL,22}^<\left(t,t\right)$ are the (1,1) and (2,2) elements of the lesser Green's function $G_{DL}^<$.
This is defined in the BdG representation as \cite{p39}
\begin{align}
G_{DL}^<(t,t)=i\sum_k\left\langle\left\langle\left(\begin{matrix}C_{k\uparrow}^\dag(t)\\C_{k\downarrow}^\dag(t)\\C_{-k\uparrow}(t)\\C_{-k\downarrow}(t)\\\end{matrix}\right)\middle|\left(\begin{matrix}d_\uparrow(t)&d_\downarrow(t)&d_\uparrow^\dag(t)&d_\downarrow^\dag(t)\\\end{matrix}\right)\right\rangle\right\rangle \label{E6}
\end{align}
The Green's function can be transformed into the energy space via a Fourier transformation: $G_{DL}^<\left(t,t\right)=\int{\frac{d\epsilon}{2\pi}}\ G_{DL}^<\left(\epsilon\right)$, and the Josephson current is written as an integral equation
\begin{eqnarray}
      I=\int d\epsilon\ i\left(\epsilon\right),
      \label{E7}
\end{eqnarray}
in which $i(\epsilon)$ is the integrand of the current in the energy space with
\begin{align}
    i\left(\epsilon\right)=\frac{e}{\pi\hbar}\left\{Re\left[\ t_sG_{DL,11}^<\left(\epsilon\right)\right]+Re\left[\ t_sG_{DL,22}^<\left(\epsilon\right)\right]\right\}.  \label{E8}
\end{align}
$G_{DL}^<\left(\epsilon\right)$ is an element of the total Green's function ${\bf G}^<\left(\epsilon\right)$. Due to the Josephson junction being in the equilibrium with zero bias, the Green's function ${\bf G}^<\left(\epsilon\right)$ can be obtained by the fluctuation-dissipation theorem
\begin{eqnarray}
    \textbf{G}^<(\varepsilon)=-f \left( \varepsilon
    \right) \left[ \textbf{G}^r\left (\varepsilon \right)-\textbf{G}^a\left( \varepsilon \right)\right], \label{E9}
\end{eqnarray}
with $f\left(\varepsilon\right)$ the Fermi distribution. $\textbf{G}^r(\varepsilon)$ and $\textbf{G}^a(\varepsilon)$ are the retarded and advanced Green's functions, and $\textbf{G}^a(\varepsilon)=[\textbf{G}^r(\varepsilon)]^\dagger$.
In addition, $\textbf{G}^r$ is defined in the space spanned by four regions: the left superconductor (L), the right superconductor (R), the central QD (D) and the magnetic impurity (M)
\begin{eqnarray}
    \textbf{G}^r=\left(\begin{matrix}G_{LL}^r&G_{LR}^r&G_{LD}^r&G_{LM}^r\\G_{RL}^r&G_{RR}^r&G_{RD}^r&G_{RM}^r\\G_{DL}^r&G_{DR}^r&G_{DD}^r&G_{DM}^r\\G_{ML}^r&G_{MR}^r&G_{MD}^r&G_{MM}^r\\\end{matrix}\right), \label{E10}
\end{eqnarray}
where each element is a $4 \times 4$ matrix in the spin space and particle-hole space under the BdG representation, similar to Eq. (\ref{E6}).
The retarded Green's function $\textbf{G}^r$ can be solved by Dyson equation
\begin{eqnarray}
    {\bf G}^r={\bf g}^r+{\bf g}^r{\bf \mathrm{\Sigma}}^r {\bf G}^r,\label{Eq11}
\end{eqnarray}
in which the self-energy ${\bf \mathrm{\Sigma}}^r$ is given by
\begin{eqnarray}
    {\bf \mathrm{\Sigma}}^r=\left(\begin{matrix}0&0&V_L&0\\0&0&V_R&0\\V_L^\dag&V_R^\dag&0&V_M^\dag\\0&0&V_M&0\\\end{matrix}\right).
\end{eqnarray}
Here $V_L$, $V_R$ and $V_M$ represent the coupling matrices of QD with left superconductor, right superconductor, and magnetic impurity, respectively. They are written as
\begin{align}
V_{L(R)}=\text{diag}(t_s,t_s,-t_s,-t_s), V_M=\text{diag}(t_c,t_c,-t_c,-t_c),
\end{align}
where $t_s$ and $t_c$ are the coupling strength in the Hamiltonian in Eq. (\ref{E4}).
$\textbf{g}^r$ in Eq.(\ref{Eq11}) is the retarded Green's function of the system without coupling between the superconductors, QD and magnetic impurity \cite{p40}
\begin{eqnarray}
    \textbf{g}^r=\text{diag}(g_{LL}^r,g_{RR}^r,g_{DD}^r,g_{MM}^r).
\end{eqnarray}
The Green's functions of the isolated superconductors are expressed as
\begin{widetext}
\begin{eqnarray}
    g^r_{LL(RR)}=-i\pi\rho\gamma(\varepsilon)\left(\begin{matrix}1&0&0&\frac{\mathrm{\Delta}}{\varepsilon}e^{i\phi_{L(R)}}\\0&1&-\frac{\mathrm{\Delta}}{\varepsilon}e^{i\phi_{L(R)}}&0\\0&-\frac{\mathrm{\Delta}}{\varepsilon}e^{-i\phi_{L(R)}}&1&0\\\frac{\mathrm{\Delta}}{\varepsilon}e^{-i\phi_{L(R)}}&0&0&1\\\end{matrix}\right),
\end{eqnarray}
\end{widetext}
where $\rho$ is the normal density of states of the superconductors, and it is a constant by using a wide-band approximation. $\gamma(\varepsilon)$ can be obtained as \cite{p41,p42}
\begin{equation}
\begin{array}{l}
\displaystyle \gamma(\varepsilon)=\frac{|\varepsilon|}{\sqrt{\varepsilon^2-\mathrm{\Delta}^2}}, |\varepsilon|>\mathrm{\Delta}, \\
\displaystyle \gamma(\varepsilon)=\frac{-i\varepsilon}{\sqrt{\mathrm{\Delta}^2-\varepsilon^2}}, |\varepsilon|<\mathrm{\Delta}.
\end{array}
\end{equation}
\hspace{\fill}

For the isolated QD and magnetic impurity, the Green's functions are solved as
\begin{widetext}
    \begin{eqnarray}
     g_{{DD}}^r=\text{diag}\left(\frac{1}{\varepsilon-\varepsilon_D+i\delta},\frac{1}{\varepsilon-\varepsilon_D+i\delta},\frac{1}{\varepsilon+\varepsilon_D+i\delta},\frac{1}{\varepsilon+\varepsilon_D+i\delta}\right),
\end{eqnarray}
\begin{eqnarray}
g_{{MM}}^r & = & \begin{pmatrix}
\frac{1}{\varepsilon-\varepsilon_M-(M_z+\alpha I)+i\delta}& 0 & 0 & 0\\
 0 & \frac{1}{\varepsilon-\varepsilon_M+(M_z+\alpha I)+i\delta} & 0 & 0\\
 0 & 0 & \frac{1}{\varepsilon+\varepsilon_M+(M_z+\alpha I)+i\delta} & 0\\
 0 & 0 & 0 &\frac{1}{\varepsilon+\varepsilon_M-(M_z+\alpha I)+i\delta}\end{pmatrix},
\end{eqnarray}
\end{widetext}
where $\delta$ is a small quantity that prevents the divergence of the Green's function.
With $\textbf{g}^r$ and ${\bf \Sigma}^r$ given, $\textbf{G}^r\left(\epsilon\right)$ is obtained and $G_{DL}^<\left(\epsilon\right)$ can be solved
through Eqs. (\ref{E9}) and (\ref{Eq11}).
Then they are substituted into Eq. (\ref{E8}) to obtain the integrand of current. By the integration in Eq. (\ref{E7}), the current is calculated.

Outside the superconducting gap, the integrand $i(\epsilon)$ of Josephson current is continuously distributed, and the current is called continuous current.
Inside the energy gap, $i(\epsilon)$ is discretely distributed around the energy of the Andreev bound states, and it is called discrete current \cite{p43,p44,p45,Zhang2013_CPR}.
The magnitude of the discrete current is approximately \cite{p47} proportional to the derivative of the energy $E_i$ of the Andreev bound state with respect to the phase: $I_{dis}\propto\sum_{i}{f\left(E_i\right)\frac{\partial E_i}{\partial\phi}}$ \cite{p44,p45,Zhang2013_CPR}. The total current consists of discrete and continuous currents, but is mainly contributed by discrete current \cite{p44,p45,Zhang2013_CPR}.
We use the linewidth function $\Gamma=2\pi\rho t_s^2$ to describe the coupling strength between the central QD and the superconductors.

In this paper, we set $\Delta=e=\hbar=1$, $\Gamma=5$ and at zero temperature,
thus the units of energy and current are chosen as $\Delta$ and $e \Delta/ \hbar$.
{The transition temperature for many conventional superconductors are several Kelvins, for example, $T_c = 7 {\rm K}$ for Pb \cite{BookSSP}, which corresponds to a superconducting gap $\Delta=1  {\rm meV}$.
Then the unit of current $e\Delta/\hbar$ is calculated to be approximately 240 nA.
All the current calculated in this paper need to be multiplied by this unit.
In our calculation results, the critical current is in the order of $0.5 e\Delta/\hbar$, corresponding to 120nA.
Although the current is not large, the present experimental techniques can sufficiently measure and distinguish it \cite{p6,p10}. }
The imaginary part of the energy is mostly taken as $\delta=0.02$.
The only exception is that when showing the distribution $i(\epsilon)$,
we choose $\delta=0.05$ to make the plot of Andreev bound states clearer [see Fig. {\ref{Fig1}(c-d)}, Fig. {\ref{Fig2}(b-d)}].

\section{\label{sec3} The appearance of SDE}

As one can see, the Josephson supercurrent $I$ is derived from the Hamiltonian, meanwhile in Eq. (\ref{E3}) the current $I$ leads to an effective Zeeman term $\widetilde{M_z}=M_z+\alpha I$ to the Hamiltonian.
The magnetic moment of the magnetic impurity is $M_z+\alpha I_0$ for the positive current $I=I_0$, while the magnetic moment is $M_z-\alpha I_0$ for the negative current $I=-I_0$.
We can qualitatively notice that the positive and negative Josephson supercurrents have different effects on the magnetic moment and Hamiltonian, resulting in the difference of the current values in the positive and negative directions after self-consistent calculation.

{Based on Ginsburg-Landau theory, the nonreciprocal current can be illustrated in a clearer way:
In a Josephson junction, the superconducting order parameter can enter from both sides of superconductors to the connecting region, where their coupling provides the possibility of the Cooper pair tunneling.
The coupling between magnetic impurity and connecting region introduces a new scattering pathway for superconducting order parameter.
Thus, the amplitude and phase of order parameter, and the effective coupling between two superconductors are regulated by the effective magnetic moment $\widetilde{M}_z=M_z+\alpha I$.
Because a positive current and a negative current lead to different values of $\widetilde{M}_z$, in turn the regulation effect from magnetic moment on superconducting coupling is different for opposite currents.
Thus, the supercurrent is nonreciprocal with $I_{c+} \ne |I_{c-}|$.
}

Next, we give a quantitative explanation.
For a given phase difference $\phi$, although the relation of the Josephson supercurrent $I$
with respect to $M_z$ is not intuitive, the relation of $I$ with respect to $\widetilde{M_z}$ is definite, see the blue curve in Fig. {\ref{Fig1}(b)}.
By combining the curve $I-\widetilde{M_z}$ with the line $\widetilde{M_z}=M_z+\alpha I$ [equivalent to $I=(\widetilde{M_z}-M_z)/\alpha$],
the current $I$($\phi$) under the given phase difference $\phi$ can be self-consistently solved. In fact, the intersection of the curve and the line
is the self-consistent result.
For a small coefficient $\alpha$, the gradient of the line $I=(\widetilde{M_z}-M_z)/\alpha$ is quite large, so there will only be one intersection with the curve $I-\widetilde{M_z}$.
This means that there will not be multiple results when calculating the current.

To explain the origin of the nonreciprocity of the Josephson supercurrent,
we compare the self-consistent results under phase differences $\phi$ and $-\phi$.
For a fixed $\widetilde{M_z}$, the currents under phase difference $\phi$ and $-\phi$ can be related through some symmetry transformation operators \cite{p48,p49}.
Here, we consider the joint transformation operator: $\mathcal{S}=\mathcal{R}U_1(-\frac{\phi}{2})$.
$U_1(-\frac{\phi}{2})$ is the $U_1$ gauge transformation with the phase $-\frac{\phi}{2}$.
$\mathcal{R}$ is the mirror reflection operator along the $x$ direction, which exchanges the left and right superconductors.
We can notice that the Hamiltonian satisfies $\mathcal{S}H(\phi)\mathcal{S}^\dag=H(-\phi)$, meanwhile the current is reversed by $\mathcal{R}$.
As a result, $I(-\phi,\widetilde{M_z})=-I(\phi,\widetilde{M_z})$ is satisfied, as shown in Fig. {\ref{Fig1}(b)}.
In addition, by using the time-reversal operator $\mathcal{T}=i\sigma_y K$
with $K$ being the complex conjugation operator,
we can derive the relation $\mathcal{T}H(\phi,\widetilde{M_z})\mathcal{T}^\dagger=H(-\phi,-\widetilde{M_z})$, and the Josephson current is reversed by $\mathcal{T}$.
As a result, the relation $I(\phi,\widetilde{M_z})=-I(-\phi,-\widetilde{M_z})$
is also satisfied, as shown in Fig. {\ref{Fig1}(b)}.
On one hand, we can see that when $M_z=0$,
the line $I=\widetilde{M_z}/\alpha$ passes through the axis origin,
and the ordinates of its intersections with the curve $I(-\phi)-\widetilde{M_z}$ and the curve $I\left(\phi\right)-\widetilde{M_z}$
must be opposite, that is, the two self-consistently calculated currents
satisfy $I\left(-\phi\right)=-I(\phi)$.
On the other hand, when $\alpha\rightarrow 0$, the line $I=(\widetilde{M_z}-M_z)/\alpha$ is parallel to the $y$ axis,
and the two self-consistently calculated currents are still opposite.
Only when both $M_z$ and $\alpha$ are nonzero,
the symmetric relation $I\left(-\phi\right)=-I(\phi)$ in conventional Josephson junctions is broken,
and $I\left(-\phi\right)\neq-I(\phi)$ can be obtained [see Fig. {\ref{Fig1}(b)}],
leading to the emergence of the SDE.

In Fig. {\ref{Fig1}(b)}, we select the curve $I-\widetilde{M_z}$ under the phase difference $\phi=\frac{\pi}{2}$ as the schematic diagram, and we can see that this curve exhibits an abrupt shift at $\left|\widetilde{M_z}\right|\approx0.3$. To explore the reason, we can plot the distribution of Andreev bound states from the discrete current $i(\epsilon)$, before and after the abrupt shift.
As shown in Fig. {\ref{Fig1}(c-d)}, the red color represents $i\left(\epsilon\right)>0$, which corresponds to a positive gradient of the Andreev bound states versus the phase $\phi$.
The blue color represents $i\left(\epsilon\right)<0$, corresponding to the negative gradient.
When $\widetilde{M_z}=0.1$, the Andreev bound states have a slight splitting due to the Zeeman effect.
At this point, although the total Josephson supercurrent is positive under $\phi=\frac{\pi}{2}$, there are two curves with positive currents and two curves with negative currents in the distribution diagram of the Andreev bound states. The total current is mostly canceled out and has a small value.
On the other hand, as shown in Fig. {\ref{Fig1}(d)},
when $\widetilde{M_z}=0.5$, the splitting of Andreev bound states becomes stronger.
At this point, one of the Andreev bound states contributing to the negative supercurrent under $\phi=\frac{\pi}{2}$ is lifted above zero energy.
According to the particle-hole symmetry, a bound state with opposite energy and opposite gradient appears and contributes to the positive supercurrent.
This results in a significant increase in the current compared to $\widetilde{M_z}=0.1$. The gradient reverses at $\left|\widetilde{M_z}\right|\approx0.3$, which corresponds to the abrupt shift in Fig. {\ref{Fig1}(b)}. The situation at $\phi=-\frac{\pi}{2}$ (or $\frac{3\pi}{2}$) is similar to that at $\phi=\frac{\pi}{2}$, except for the opposite sign of the current.

\begin{figure}[!htb]
\centerline{\includegraphics[width=\columnwidth]{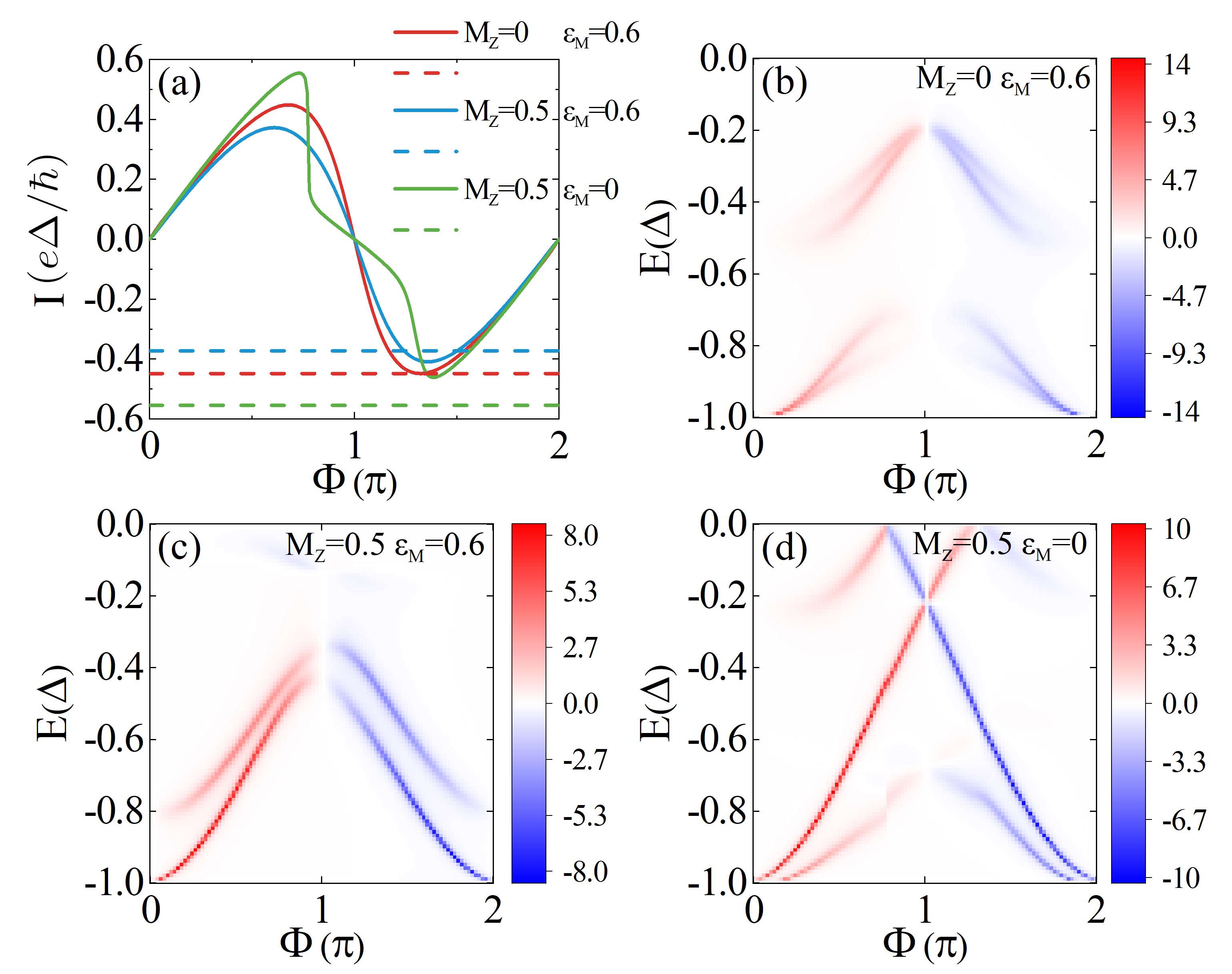}}
\caption{(a) The Josephson current $I$-superconducting phase difference $\phi$ relations for various parameters are represented by the solid lines.
The dashed horizontal lines are $I=-I_{c+}$, with $I_{c+}$ the maxima of the current $I$. (b-d) The energy distribution of the integrand $i(\epsilon)$ of the current under the corresponding parameters in (a). Other parameters are set as $\epsilon_D=0$, $t_c=1$, $\alpha=0.2$.
}\label{Fig2}
\end{figure}

Next, we demonstrate the self-consistently calculated Josephson current $I-\phi$ relation in Fig. {\ref{Fig2}(a)}.
The positive and negative critical currents of the Josephson junction correspond to the maximum and minimum values of the curve $I(\phi)$, and are indicated as $I_{c+}$ and $I_{c-}$, respectively.
The phenomenon of the unequal critical currents in opposite directions $I_{c+}\neq\left|I_{c-}\right|$ is called the SDE.
In Fig. {\ref{Fig2}(a)}, we extract the maximum value $I_{c+}$ of each curve
and plot the dashed horizontal lines to illustrate $I=-I_{c+}$.
When $M_z=0$, the horizontal line $I=-I_{c+}$ is exactly tangent
to the curve $I(\phi)$, that is, $I_{c+}=\left|I_{c-}\right|$.
Also, the curve $I(\phi)$ satisfies the relation $I\left(-\phi\right)=-I(\phi)$
at $M_z=0$, as the above discussion in Fig. {\ref{Fig1}(b)}.
On the contrary, when $M_z\neq0$, the symmetry of the current, $I\left(-\phi\right)=-I(\phi)$, is broken,
leading to the relation $I_{c+}\neq\left|I_{c-}\right|$, and the SDE emerges.
In addition, the current satisfies $I=0$ at $\phi=0$, $\pi$.
This is because the relations $I(-\phi,\widetilde{M_z})=-I(\phi,\widetilde{M_z})$
and $I(\phi,\widetilde{M_z})=I(\phi+2\pi,\widetilde{M_z})$ ensure $I(\widetilde{M_z})=0$ at $\phi=0$, $\pi$, and the self-consistently solved current must be zero.
This indicates that our proposal is a conventional Josephson junction, instead of a $\phi_0$ junction \cite{p7,p8,p17,p20,p21,p22,p23,p24,p25,p26,p28,p29,p31,p48,p49,p50,p51,p52,p53}. It is quite different from previous studies, where the SDE usually demands a $\phi_0$ junction with $I\left(\phi=0\right)\neq0$ \cite{p7,p8,p17,p20,p21,p22,p23,p24,p25,p26,p28,p29,p31}.

In Fig. {\ref{Fig2}(b-d)}, we plot the distribution of the integrand $i(\epsilon)$ of the supercurrent under the three sets of parameters in Fig. {\ref{Fig2}(a)} to illustrate the corresponding Andreev bound states.
In Fig. {\ref{Fig2}(b)}, the distribution of Andreev bound states with $M_z=0$
is symmetric about $\phi=\pi$, which indicates the relation $I\left(-\phi\right)=-I(\phi)$ \cite{p44}.
In Figs. {\ref{Fig2}(c) and \ref{Fig2}(d)},
due to the different effects of the positive
and negative currents on the magnetic moment with $M_z\neq0$,
the distribution of Andreev bound states become asymmetric, corresponding to the relation $I\left(-\phi\right)\neq-I(\phi)$.
By adjusting the energy level of the magnetic impurity, the Andreev bound state can cross zero energy and appear in the positive-energy region (see Fig. {\ref{Fig2}(d)}).
Due to the particle-hole symmetry, a bound state with opposite gradient will appear in the negative-energy region, leading to the result that the current carried by the state changes its sign and the total current shifts abruptly, as shown by the green line in Fig. {\ref{Fig2}(a)}.
The current shifts abruptly at $\phi\approx0.75\pi$, $1.3\pi$, and the shift phases are also asymmetric about $\phi=\pi$.
This makes the difference of positive and negative critical currents more remarkable.

\section{\label{sec4}The regulation on SDE}

In order to study the SDE in detail, here we introduce $\eta$ to represent the superconducting diode efficiency, which is defined as
\begin{eqnarray}
    \eta=\frac{I_{c+}-|I_{c-}|}{(I_{c+}+|I_{c-}|)/2}.
\end{eqnarray}

\begin{figure}[!htb]
\centerline{\includegraphics[width=\columnwidth]{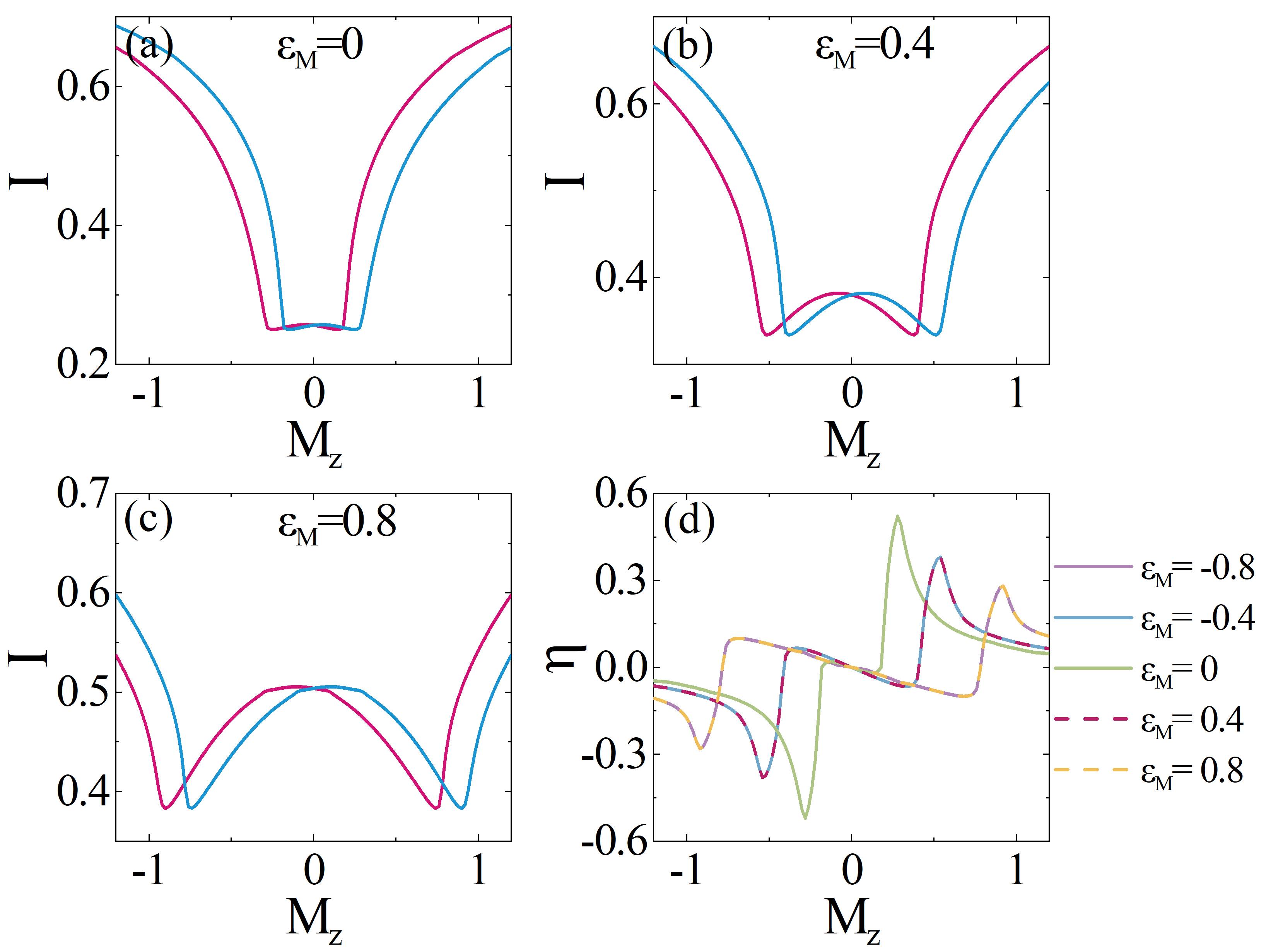}}
\caption{(a-c) Positive and negative critical currents $I_{c+}$ (red) and $\left|I_{c-}\right|$ (blue) versus intrinsic magnetic moment $M_z$ for various $\epsilon_M$. (d) Superconducting diode efficiency $\eta$ versus $M_z$. Parameters: $\epsilon_D=0$, $t_c=1$, $\alpha=0.2$.}\label{Fig3}
\end{figure}

In Fig. {\ref{Fig3}(a-c)}, we demonstrate the positive and negative critical currents $I_{c+}$ and $\left|I_{c-}\right|$ versus the intrinsic magnetic moment $M_z$ with different parameters $\epsilon_M=0, 0.4, 0.8$.
These two curves have similar shapes and are symmetric about $M_z=0$.
Meanwhile, the diode efficiency $\eta$ versus $M_z$ with different parameters $\epsilon_M$ is shown in Fig. {\ref{Fig3}(d)}.
The diode efficiency $\eta$ is an odd function with respect to $M_z$,
and its value can reach the maximum and shows a peak under specific $M_z$.
When $\epsilon_M$ increases from 0, the extremum of the curve $\eta-M_z$ decreases, and the corresponding value of $|M_z|$ at the extremum becomes larger. Moreover, the curves $\eta-M_z$ at the energy levels $\epsilon_M$ and $-\epsilon_M$ completely coincide.
In the following, we explain the origin of these results.

By the spin rotation operator $U_s=e^{i\frac{\pi}{2}\sigma_x}$,
the Hamiltonian is transformed as $U_s H\left(M_z,I\right)U_s^\dag=H\left(-M_z,-I\right)$.
This means that when the current and magnetic moment are simultaneously reversed, the corresponding Hamiltonians are equivalent.
Therefore, the relation between the positive and negative critical currents satisfies $I_{c+}\left(M_z\right)=\left|I_{c-}(-M_z)\right|$. As a result, the efficiency $\eta$ is an odd function of $M_z$ in Fig. {\ref{Fig3}(d)}
\begin{eqnarray}
    \eta\left(-M_z\right)=-\eta\left(M_z\right).  \label{E20}
\end{eqnarray}

The effect of current and intrinsic magnetic moment on our calculated results
is directly reflected in the value of $\widetilde{M_z}=M_z+\alpha I$.
So, the curve of the positive critical current $I_{c+}-M_z$ can be
approximately regarded as the result of left translation of the curve
$I-\widetilde{M_z}$ in Fig. {\ref{Fig1}(b)}.
The shapes of curves in Fig. {\ref{Fig3}(a)} and Fig. {\ref{Fig1}(b)} seem different, because the ranges of the vertical coordinate are different.
Similarly, the curve $\left|I_{c-}\right|-M_z$ can be regarded as the result of the right translation of the curve $I-\widetilde{M_z}$ in Fig. {\ref{Fig1}(b)}.
The dislocation of the two curves will lead to the result that $I_{c+}\left(M_z\right)\neq\left|I_{c-}(M_z)\right|$, and the SDE occurs.
A larger gradient of the curve $I-\widetilde{M_z}$ corresponds to a greater difference between $I_{c+}$ and $\left|I_{c-}\right|$ at the corresponding magnetic moment $M_z$.
As shown by the analysis in Sec. \ref{sec3}, the curve $I-\widetilde{M_z}$
has an abrupt shift with a large gradient at specific magnetic moments due to the evolution of the Andreev bound states.
So, at such value of magnetic moment, the difference between the positive and negative critical currents is the most significant, and the efficiency will achieve the maximum, which corresponds to the peak in Fig. {\ref{Fig3}(d)}.
When the energy level $\epsilon_M$ increases from zero,
the shift amplitude of $I-\widetilde{M_z}$ becomes smaller,
and the extremum of efficiency $\eta$ decreases.

Next, we consider a joint symmetry operator $\mathcal{W}=\mathcal{RTP}U_1\left(\frac{\pi+\phi}{2}\right)$, where $\mathcal{R}$ is the mirror reflection operator along the $x$ direction, $\mathcal{T}=i\sigma_y K$ is the time-reversal operator, $\mathcal{P}=\tau_x K$ is the particle-hole symmetry operator and $U_1\left(\frac{\pi+\phi}{2}\right)$ is the $U_1$ gauge transformation with the phase $\frac{\pi+\phi}{2}$, with $\tau_x$ the Pauli matrix in the particle-hole space.
Under the operations of these transformations, the current remains unchanged,
the Hamiltonians of the left and right superconductors remain unchanged, and magnetic moment $M_z$ also remains unchanged, only the energy level term of QD and magnetic impurity reverses the sign.
That is, the Hamiltonian satisfies $\mathcal{W}H\left(\epsilon_D,\epsilon_M\right)\mathcal{W}^\dag=H\left(-\epsilon_D,-\epsilon_M\right)$. As a result, the current satisfies $I_{c+\left(-\right)}\left(\epsilon_D,\epsilon_M\right)=I_{c+\left(-\right)}\left(-\epsilon_D,-\epsilon_M\right)$, and the efficiency satisfies
\begin{eqnarray}
    \eta\left(\epsilon_D,\epsilon_M\right)=\eta\left({-\epsilon_D,-\epsilon_M}\right). \label{E21}
\end{eqnarray}
So, we can see that when $\epsilon_D=0$, the curves $\eta\left(\epsilon_M\right)-M_z$ and $\eta\left(-\epsilon_M\right)-M_z$ completely coincide in Fig. {\ref{Fig3}(d)}.

\begin{figure}[!htb]

\centerline{\includegraphics[width=\columnwidth]{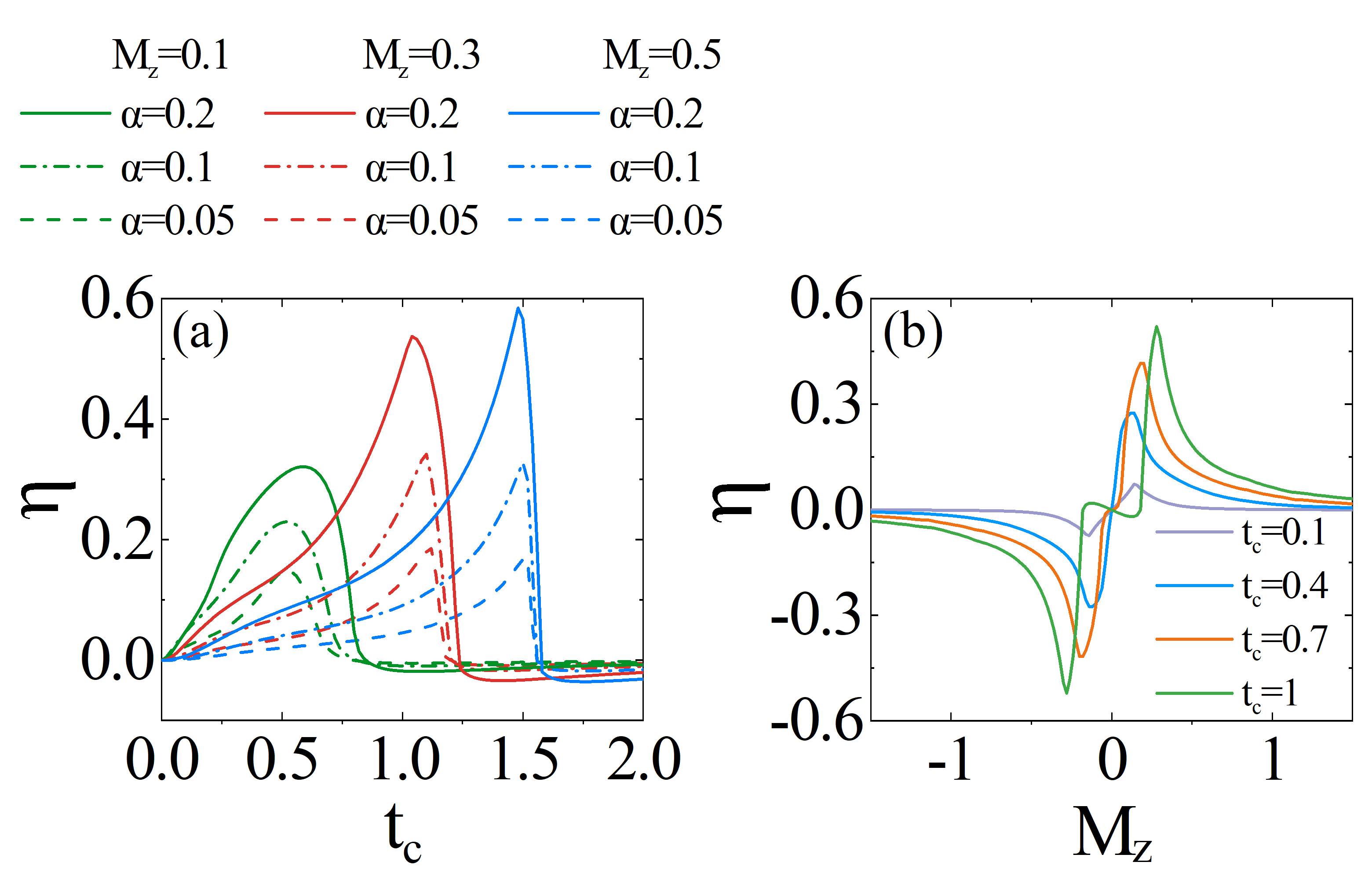}}
\caption{(a) The efficiency $\eta$ versus the coupling strength $t_c$ with various intrinsic magnetic moment $M_z$ and proportional coefficient $\alpha$.
(b) The efficiency $\eta$ versus the magnetic moment $M_z$ with various $t_c$ at $\alpha=0.2$. Other parameters: $\epsilon_D=\epsilon_M=0$.}\label{Fig4}
\end{figure}

In Fig. {\ref{Fig4}(a)}, we plot the superconducting diode efficiency $\eta$ versus the coupling strength of QD-magnetic impurity $t_c$ at $M_z=0.1, 0.3, 0.5$ in green, red and blue curves.
And for each $M_z$, we distinguish the cases at $\alpha=0.2, 0.1, 0.05$ by using the solid line, dash-dotted line and dashed line.
When the magnetic moment $M_z$ is fixed, the larger $\alpha$ corresponds to the greater effect of the current on the magnetic moment, and the obtained efficiency are higher.
We can see that the efficiency $\eta$ increases from zero as the coupling strength $t_c$ increases from 0.
That is because the SDE results from the impacts of the magnetic impurity on the Josephson junction.
Only when there is a coupling between the magnetic impurity and the connection region in the Josephson junction ($t_c\neq0$), the SDE can emerge.
However, when $t_c$ is too large, the effect of scattering caused by the magnetic impurity plays a key role in the system, and the current will decrease.
At this point, the modulation on the magnetic moment by the current is severely weakened, causing a decrease in $\eta$.

In Fig. {\ref{Fig4}(b)}, we show the superconducting diode efficiency $\eta$ versus the magnetic moment $M_z$ with different $t_c$, and these curves still satisfy the properties of the odd function in Eq. (\ref{E20}).
Simultaneously, similar to Fig. {\ref{Fig3}(d)}, the efficiency $\eta$ will achieve the extremum value at a specific $M_z$ as a result of the evolution of the Andreev bound states. Here, $t_c$ will slightly affect the distribution of the Andreev bound states, making a slight impact on $M_z$ that corresponds to the peak of $\eta$.

\begin{figure}[!htb]
\centerline{\includegraphics[width=\columnwidth]{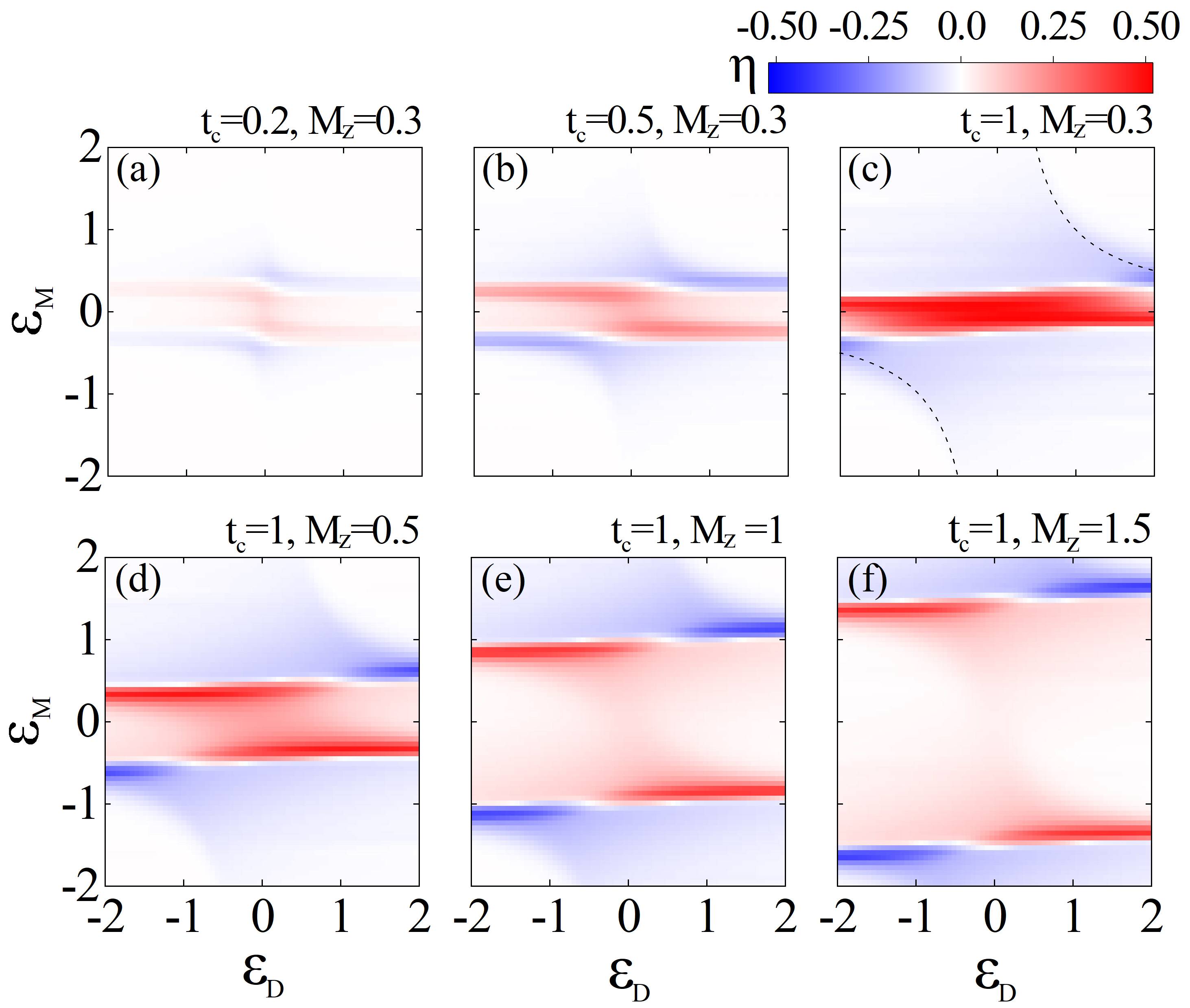}}
\caption{The superconducting diode efficiency $\eta$ versus $\varepsilon_D$ and $\varepsilon_M$ with $\alpha=0.2$. (a-c) $M_z=0.3$ for different $t_c$. (c-f) $t_c=1$ for different $M_z$.
The black dotted curve in (c) demonstrates the hyperbola $\varepsilon_D \varepsilon_M =t_c^2$. }\label{Fig5}
\end{figure}

To comprehensively study the modulation on SDE by various parameters, we illustrate the plots of the efficiency $\eta$ in the $\left(\varepsilon_D,\varepsilon_M\right)$ space under different coupling strength $t_c$ and magnetic moment $M_z$ in Fig. {\ref{Fig5}}.
As for a single plot [e.g. Fig. {\ref{Fig5}(c)}],
when adjusting the energy levels of the central QD $\varepsilon_D$ and the magnetic impurity $\varepsilon_M$,
the efficiency will be sightly impacted by $\varepsilon_D$ but greatly impacted by $\varepsilon_M$.
That is because the emergence of the SDE derives from the magnetic moment $M_z$, which directly acts and regulates the effective energy level of magnetic impurity $\varepsilon_M$, making the impact of $\varepsilon_M$ more significant than that of $\varepsilon_D$.
When $\epsilon_M$ is adjusted, there will be an abrupt shift in $\eta$,
changing its sign from positive (red) to negative (blue).

In Fig. {\ref{Fig5}(a-c)}, when $M_z$ is fixed at 0.3, the maximum of the efficiency $\eta$ can still reach about 0.1 even under the weak coupling strength $t_c=0.2$.
As $t_c$ increases, the overall SDE will become more significant:
the efficiency $\eta$ can even exceed 0.5 at $t_c=1$.
As $M_z$ increases from Fig. {\ref{Fig5}(c)} to Fig. {\ref{Fig5}(f)}, $\left|\epsilon_M\right|$ corresponding to the abruptly shifted $\eta$ increases as well, and the parameter space with positive values of $\eta$ becomes wider.
Besides, the symmetry relation Eq. (\ref{E21}) derived by the joint symmetry operator $\mathcal{W}$ is intuitively apparent in all the pictures in Fig. {\ref{Fig5}}.

In Fig. {\ref{Fig5}(c)}, the value of efficiency $\eta$ goes through a sudden drop when the parameters approximately satisfy $\varepsilon_D\varepsilon_M=t_c^2$,
and $\eta$ is very small at $\varepsilon_D\varepsilon_M >t_c^2$,
which is reflected in the form of hyperbola in the picture
[see the black dotted curves in Fig. {\ref{Fig5}(c)}].
In this regard, Ref. \cite{p42} investigated the Josephson junction
at $M_z=0$ and $\alpha=0$:
It was found that when $\varepsilon_D\varepsilon_M>t_c^2$, the current obviously reduces due to the evolution of the Andreev bound states.
Since in Fig. \ref{Fig5}(c) the value of $\alpha$ is small and $M_z$ is much smaller than  $t_c$,
the results in Ref. \cite{p42} are approximately obeyed in our system.
And due to the reduction in current, the impact on the magnetic moment is weakened, and the efficiency decreases.

At last, we estimate the magnitude of the coefficient $\alpha$.
We suppose that there are some magnetic atoms forming an impurity cluster with $R=1 {\rm nm}$ away from the connection region of the Josephson junction.
The Zeeman energy is $\frac{1}{2}g\mu_B B$, where $g$ is the Lande factor, $\mu_B$ is the Bohr magneton and $B=\mu_0\left(1+\chi\right)H$ is the induced magnetic field with the vacuum permeability $\mu_0$ and the magnetic susceptibility $\chi$.
According to the Ampere circuital theorem, one can obtain $H=I/2\pi R$.
As a result, the Zeeman energy caused by the current can be expressed as $\alpha I=g\mu_B\mu_0I\left(1+\chi\right)/4\pi R$, and $\alpha=g\mu_B\mu_0\left(1+\chi\right)/4\pi R$.
Note that the units of energy and current are $\Delta$ and $e\Delta/\hbar$, hence the unit of $\alpha$ is $\hbar/e$.
Due to the exchange interactions, the magnetic moment in ferromagnetic systems can go through a significant change by merely varying a very small order of magnitude of $H$ \cite{p54,p55,Ding2021_chl}, indicating that the magnetic susceptibility $\chi$ has a considerable value.
In fact, the susceptibility $\chi$ of the ferromagnetic materials
can even reach $10^6$ orders of magnitude \cite{p56}.
Here we just set $\chi=20000$, and when $g=2$, we can obtain $\alpha=0.06$.
Meanwhile, if the impurity can have a higher Lande factor, the coefficient $\alpha$ can be larger.
In summary, it is reasonable for us to set the magnitude of $\alpha$ at $10^{-2}$ and $10^{-1}$, and this magnitude of $\alpha$ can already achieve significant SDEs in the wide parameter space (see Fig. {\ref{Fig4}}(a)): Even at $\alpha=0.05$, the diode efficiency $\eta$ can reach 0.3; At $\alpha=0.2$, the efficiency $\eta$ can even reach 0.5.

\section{\label{sec5}Discussion and Conclusion}

{In usual, besides the current-phase relation $I(\phi)$,
a Josephson junction also has the resistance $R$ and capacitance $C$
and can be described by the resistively and capacitively
shunted junction (RCSJ) model \cite{BookITS}.
Under a voltage bias $V$ between the junction, we have
$ I_C = C dV/ dt $ and $ I_R = V / R $.
Due to $V = \frac{\hbar}{2e} \frac{d\phi}{dt} $, the total current
through the Josephson junction is $I_t = I(\phi) +I_C +I_R$, and
the total current-phase relation can be expressed as:
\begin{equation}
	\frac{C \hbar}{2e} \frac{d^2 \phi}{dt^2} + \frac{\hbar}{2eR} \frac{d\phi}{dt} + I(\phi)-I_t =0. \label{RCSJ}
\end{equation}
By a substitution $\phi \rightarrow$ coordinate $x$,
one can find that the phase evolution in Eq.(\ref{RCSJ}) is equivalent to
a mechanical equation: It describes the motion of item
with mass $\frac{C \hbar}{2e}$ under a potential $U(\phi)$ and
viscous drag force $\frac{\hbar}{2eR} \frac{d\phi}{dt}$.
The potential is called washboard potential
satisfying $dU(\phi)/d\phi = I(\phi) - I_t$.
Under a total current $I_t$, when the item keeps stationary with $dx/dt=0$ (corresponding to $d\phi/dt=0$), the voltage is $V= \frac{\hbar}{2e} \frac{d\phi}{dt} =0$.
This means that the current is a supercurrent instead
of a dissipative current.
In our Josephson junction, the magnetic impurity causes the peculiar $I(\phi)$ relation with unequal critical currents in opposite directions $I_{c+} \ne |I_{c-}|$.
When $I_t = 0$, the schematic plot of the corresponding
washboard potential $U(\phi)$ is shown as the gray curve in Fig. \ref{Fig6}.
Because the relation $I(-\phi) = -I(\phi)$ is broken,
$U(\phi)$ becomes asymmetric about $\phi = n \pi$ ($n$ is integer).
Also, note that when $I_t = 0$, $dU/d\phi = I(\phi)$, thus the relation
$I_{c+} \ne |I_{c-}|$ results in $(dU/d\phi)_{max}\ne |(dU/d\phi)_{min}|$.
When under a large total current $|I_t|$, it can sharply tilt the
washboard potential $U(\phi)$, so that the stable points disappear
and the current becomes dissipative.
Because of the asymmetry of the gradient of $U(\phi)$,
the critical $I_t$ that destroys stability is $I_{t,c+}$ in positive direction
and $I_{t,c-}$ in negative direction, with different absolute values.
Fig. \ref{Fig6} shows a case that $I_{t,c+} > |I_{t,c-}|$, $(dU/d\phi)_{max} > |(dU/d\phi)_{min}|$.
For the positive $I_t$ with $|I_{t,c-}|<I_t<I_{t,c+}$,
the stable points exist in the washboard potential $U(\phi)$
[see the blue curve in Fig. \ref{Fig6}],
but for the negative $-I_t$, there is no stable point in $U(\phi)$
[see the red curve in Fig. \ref{Fig6}].
So the nonreciprocal critical current with $I_{t,c+} \ne |I_{t,c-}|$
can still exist even if the Josephson junction has the resistance $R$ and capacitance $C$.

\begin{figure}[!htb]
	\centerline{\includegraphics[width=0.5\columnwidth]{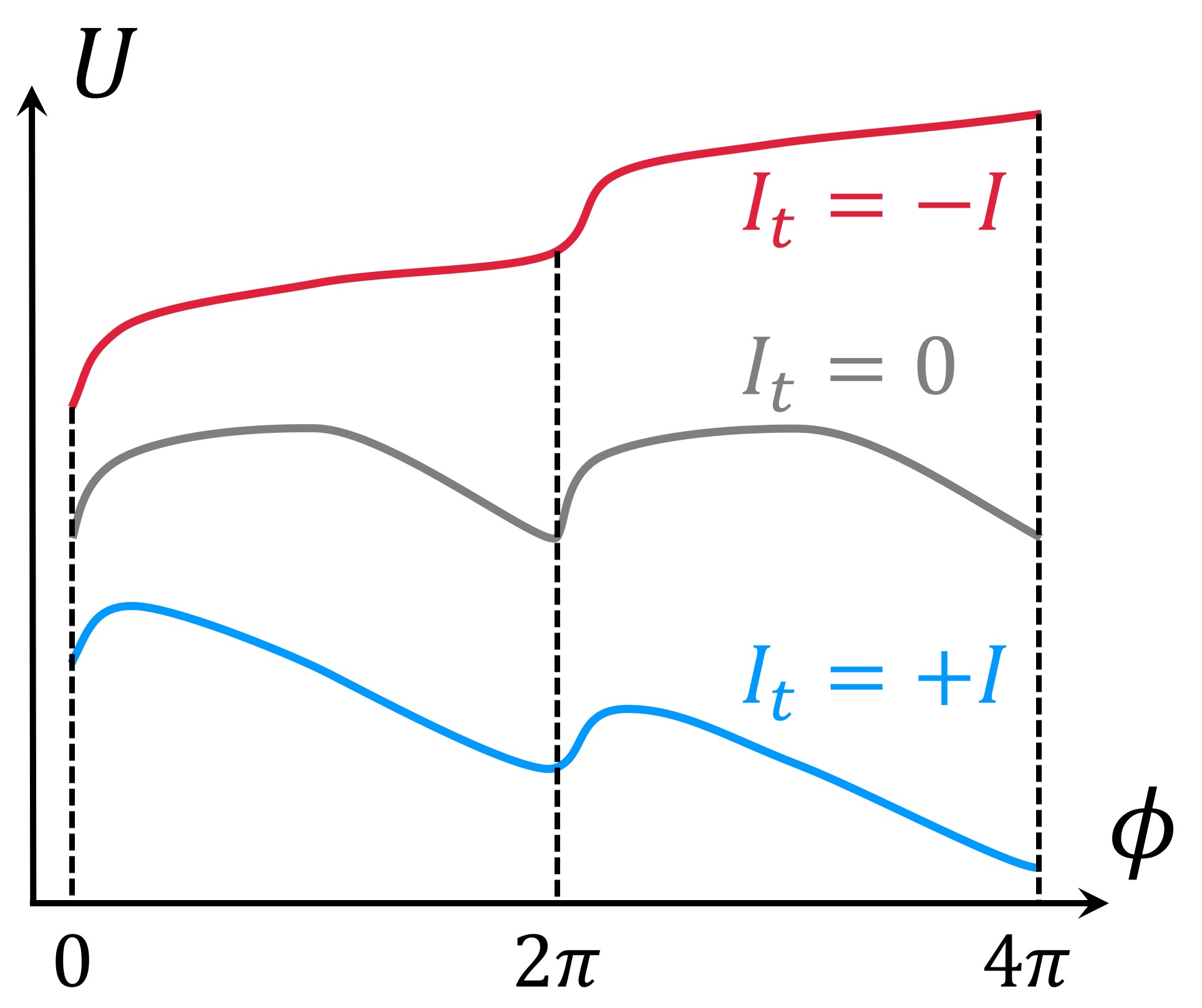}}
	\caption{{Schematic plots for washboard potential $U(\phi)$
of Josephson junction coupled to magnetic impurity.
	The gray, red, blue curves respectively correspond to zero, negative,
and positive total current $I_t$.} }\label{Fig6}
\end{figure}

In addition, our Josephson diodes based on magnetic impurities
can be applied to construct superconducting interferometers.
For example, let us consider that a superconducting interferometer
consists of two Josephson junction (an our Josephson junction
and a conventional Josephson junction) connected in parallel and
threaded by a magnetic flux.
The absolute values of the critical current of this
superconducting interferometer in positive and negative
directions are still unequal. }

A very recent experiment \cite{p10} fabricated a Josephson junction
by a magnetic atom connected with two superconductors,
and observed the result of the unequal critical currents
in positive and negative directions.
In fact, the experimental results can be explained
by the mechanism in our model.
In the experiment, it is quite hard to make the magnetic atom accurately
located at the center of Josephson junction,
and the current will unavoidably deviate from the center of the magnetic atom.
As a result, a magnetic field should be induced
and affect the magnetic moment of the atom.
Therefore, our theory is applicable to this experiment.
Although there is only a single magnetic atom with a relatively low magnetic susceptibility $\chi$, the distance $R$ is much smaller than $1 \rm nm$,
thus the coefficient $\alpha$ is not quite small.

In summary, we propose a simplified and universal Josephson diode device: a Josephson junction with its connection region coupled with a magnetic impurity.
In this device, the SDE is induced by the different impacts of positive and negative supercurrents on magnetic impurity.
What's more, the symmetry relations of supercurrents are derived
and the effect of the system's parameters on diode efficiency is investigated.
We find the remarkable SDE in the wide parameter space.
The SDE efficiency can be regulated by controlling gate voltage (to change energy level) and moving the magnetic impurity closer to the connection region (to change $t_c$).
In particular, our mechanism can be applied to any kind of superconductor, including conventional $s$-wave superconductors, and it does not rely on the unconventional finite-momentum Cooper pair property. Besides, our proposal only demands a magnetic impurity, while the spin-orbit coupling, chirality and the external magnetic field are not required.
This universal proposal provides a new theoretical perspective
for the further development of dissipationless electronic devices.

\begin{acknowledgments}
We thank Yu-Chen Zhuang for fruitful discussions.
This work was financially supported by the National Natural
Science Foundation of China (Grant No. 12374034 and No. 11921005), the Innovation Program for Quantum Science and Technology
(2021ZD0302403), and the Strategic priority Research
Program of Chinese Academy of Sciences (Grant No.
XDB28000000).
\end{acknowledgments}


\begin{thebibliography}{30}
\bibitem{p1}
F. Ando, Y. Miyasaka, T. Li, J. Ishizuka, T. Arakawa, Y. Shiota, T. Moriyama, Y. Yanase, and T. Ono,
\href{https://www.nature.com/articles/s41586-020-2590-4}{Nature (London) \textbf{584}, 373 (2020)}.
\bibitem{p2}
L. Bauriedl, C. B{\" a}uml, L. Fuchs, C. Baumgartner, N. Paulik, J. M. Bauer, K.-Q. Lin, J. M. Lupton, T. Taniguchi, and K. Watanabe, C. Strunk, and N. Paradiso,
\href{https://www.nature.com/articles/s41467-022-31954-5}{Nat. Commun. \textbf{13}, 4266 (2022)}.
\bibitem{p3}
H. Narita, J. Ishizuka, R. Kawarazaki, D. Kan, Y. Shiota, T. Moriyama, Y. Shimakawa, A. V. Ognev, A. S. Samardak, Y. Yanase, and T. Ono,
\href{https://www.nature.com/articles/s41565-022-01159-4}{Nat. Nanotechnol. \textbf{17}, 823 (2022)}.
\bibitem{p4}
J.-X. Lin, P. Siriviboon, H. D. Scammell, S. Liu, D. Rhodes, K. Watanabe, T. Taniguchi, J. Hone, M. S. Scheurer, and J. I. A. Li,
\href{https://www.nature.com/articles/s41567-022-01700-1}{Nat. Phys. \textbf{18}, 1221 (2022)}.
\bibitem{p5}
B. Pal, A. Chakraborty, P. K. Sivakumar, M. Davydova, A. K. Gopi, A. K. Pandeya, J. A. Krieger, Y. Zhang, M. Date, S. Ju, N. Yuan, N. B. M. Schr{\" o}ter, L. Fu, and S. S. P. Parkin,
\href{https://www.nature.com/articles/s41567-022-01699-5}{Nat. Phys. \textbf{18}, 1228 (2022)}.
\bibitem{p6}
B. Turini, S. Salimian, M. Carrega, A. Iorio, E. Strambini, F. Giazotto, V. Zannier, L. Sorba, and S. Heun,
\href{https://pubs.acs.org/doi/full/10.1021/acs.nanolett.2c02899}{Nano Letters \textbf{22}, 8502 (2022)}.
\bibitem{p7}
K.-R. Jeon, J.-K. Kim, J. Yeon, J.-C. Jeon, H. Han, A. Cottet, T. Kontos, and S. S. Parkin,
\href{https://www.nature.com/articles/s41563-022-01300-7}{Nat. Mater. \textbf{21}, 1008 (2022)}.
\bibitem{p8}
C. Baumgartner, L. Fuchs, A. Costa, S. Reinhardt, S. Gronin, G. C. Gardner, T. Lindemann, M. J. Manfra, P. E. Faria Junior, D. Kochan, J. Fabian, N. Paradiso, and C. Strunk,
\href{https://www.nature.com/articles/s41565-021-01009-9}{Nat. Nanotechnol. \textbf{17}, 39 (2022)}.
\bibitem{p9}
H. Wu, Y. Wang, Y. Xu, P. K. Sivakumar, C. Pasco, U. Filippozzi, S. S. P. Parkin, Y.-J. Zeng, T. McQueen, and M. N. Ali,
\href{https://www.nature.com/articles/s41586-022-04504-8}{Nature (London) \textbf{604}, 653 (2022)}.
\bibitem{p10}
M. Trahms, L. Melischek, J. F. Steiner, B. Mahendru, I. Tamir, N. Bogdanoff, O. Peters, G. Reecht, C. B. Winkelmann, F. von Oppen, and K. J. Franke,
\href{https://www.nature.com/articles/s41586-023-05743-z}{Nature (London) \textbf{615}, 628 (2023)}.
\bibitem{p11}
A. Daido, Y. Ikeda, and Y. Yanase,
\href{https://journals.aps.org/prl/abstract/10.1103/PhysRevLett.128.037001}{Phys. Rev. Lett. \textbf{128}, 037001 (2022)}.
\bibitem{p12}
S. Ili{\' c} and F. S. Bergeret,
\href{https://journals.aps.org/prl/abstract/10.1103/PhysRevLett.128.177001}{Phys. Rev. Lett. \textbf{128}, 177001 (2022)}.
\bibitem{p13}
N. F. Q. Yuan and L. Fu,
\href{https://www.pnas.org/doi/abs/10.1073/pnas.2119548119}{Proc. Natl. Acad. Sci. U.S.A. \textbf{119}, e2119548119 (2022)}.
\bibitem{p14}
J. J. He, Y. Tanaka, and N. Nagaosa,
\href{https://iopscience.iop.org/article/10.1088/1367-2630/ac6766/meta}{New Journal of Physics \textbf{24}, 053014 (2022)}.
\bibitem{p15}
H. F. Legg, D. Loss, and J. Klinovaja,
\href{https://journals.aps.org/prb/abstract/10.1103/PhysRevB.106.104501}{Phys. Rev. B \textbf{106}, 104501 (2022)}.
\bibitem{p16}
Y. Mao, Q. Yan, Y.-C. Zhuang, and Q.-F. Sun,
\href{https://arxiv.org/abs/2306.09113}{arxiv: 2306.09113}.
\bibitem{p17}
M. Davydova, S. Prembabu, and L. Fu,
\href{https://www.science.org/doi/full/10.1126/sciadv.abo0309}{Science Advances \textbf{8}, eabo0309 (2022)}.
\bibitem{p18}
J. Hu, C. Wu, and X. Dai,
\href{https://journals.aps.org/prl/abstract/10.1103/PhysRevLett.99.067004}{Phys. Rev. Lett. \textbf{99}, 067004 (2007)}.
\bibitem{p19}
Y. Zhang, Y. Gu, P. Li, J. Hu, and K. Jiang,
\href{https://journals.aps.org/prx/abstract/10.1103/PhysRevX.12.041013}{Phys. Rev. X \textbf{12}, 041013 (2022)}.
\bibitem{p20}
A. Zazunov, R. Egger, T. Jonckheere, and T. Martin,
\href{https://journals.aps.org/prl/abstract/10.1103/PhysRevLett.103.147004}{Phys. Rev. Lett. \textbf{103}, 147004 (2009)}.
\bibitem{p21}
A. Brunetti, A. Zazunov, A. Kundu, and R. Egger,
\href{https://journals.aps.org/prb/abstract/10.1103/PhysRevB.88.144515}{Phys. Rev. B \textbf{88}, 144515 (2013)}.
\bibitem{p22}
T. Yokoyama, M. Eto, and Y. V. Nazarov,
\href{https://journals.aps.org/prb/abstract/10.1103/PhysRevB.89.195407}{Phys. Rev. B \textbf{89}, 195407 (2014)}.
\bibitem{p23}
A. Buzdin,
\href{https://journals.aps.org/prl/abstract/10.1103/PhysRevLett.101.107005}{Phys. Rev. Lett. \textbf{101}, 107005 (2008)}.
\bibitem{p24}
F. Dolcini, M. Houzet, and J. S. Meyer,
\href{https://journals.aps.org/prb/abstract/10.1103/PhysRevB.92.035428}{Phys. Rev. B \textbf{92}, 035428 (2015)}.
\bibitem{p25}
A. A. Reynoso, G. Usaj, C. A. Balseiro, D. Feinberg, and M. Avignon,
\href{https://journals.aps.org/prl/abstract/10.1103/PhysRevLett.101.107001}{Phys. Rev. Lett. \textbf{101}, 107001 (2008)}.
\bibitem{p26}
A. A. Kopasov, A. G. Kutlin, and A. S. Mel'nikov,
\href{https://journals.aps.org/prb/abstract/10.1103/PhysRevB.103.144520}{Phys. Rev. B \textbf{103}, 144520 (2021)}.
\bibitem{p27}
K. Halterman, M. Alidoust, R. Smith, and S. Starr,
\href{https://journals.aps.org/prb/abstract/10.1103/PhysRevB.105.104508}{Phys. Rev. B \textbf{105}, 104508 (2022)}.
\bibitem{p28}
Y. V. Fominov and D. S. Mikhailov,
\href{https://journals.aps.org/prb/abstract/10.1103/PhysRevB.106.134514}{Phys. Rev. B \textbf{106}, 134514 (2022)}.
\bibitem{p29}
R. S. Souto, M. Leijnse, and C. Schrade,
\href{https://journals.aps.org/prl/abstract/10.1103/PhysRevLett.129.267702}{Phys. Rev. Lett. \textbf{129}, 267702 (2022)}.
\bibitem{p30}
Q. Cheng and Q.-F. Sun,
\href{https://journals.aps.org/prb/abstract/10.1103/PhysRevB.107.184511}{Phys. Rev. B \textbf{107}, 184511 (2023)}.
\bibitem{p31}
J.-X. Hu, Z.-T. Sun, Y.-M. Xie, and K. T. Law,
\href{https://journals.aps.org/prl/abstract/10.1103/PhysRevLett.130.266003}{Phys. Rev. Lett. \textbf{130}, 266003 (2023)}.
\bibitem{p32}
J. F. Steiner, L. Melischek, M. Trahms, K. J. Franke, and F. von Oppen,
\href{https://journals.aps.org/prl/abstract/10.1103/PhysRevLett.130.177002}{Phys. Rev. Lett. \textbf{130}, 177002 (2023)}.
\bibitem{p33}
G. L. J. A. Rikken, J. F{\" o}lling, and P. Wyder,
\href{https://journals.aps.org/prl/abstract/10.1103/PhysRevLett.87.236602}{Phys. Rev. Lett. \textbf{87}, 236602 (2001)}.
\bibitem{p34}
T. Morimoto and N. Nagaosa,
\href{https://journals.aps.org/prl/abstract/10.1103/PhysRevLett.117.146603}{Phys. Rev. Lett. \textbf{117}, 146603 (2016)}.
\bibitem{p35}
R. Wakatsuki, Y. Saito, S. Hoshino, Y. M. Itahashi, T. Ideue, M. Ezawa, Y. Iwasa, and N. Nagaosa,
\href{https://www.science.org/doi/full/10.1126/sciadv.1602390}{Science Advances \textbf{3}, e1602390 (2017)}.
\bibitem{p36}
R. Wakatsuki and N. Nagaosa,
\href{https://journals.aps.org/prl/abstract/10.1103/PhysRevLett.121.026601}{Phys. Rev. Lett. \textbf{121}, 026601 (2018)}.
\bibitem{p37}
S. Hoshino, R. Wakatsuki, K. Hamamoto, and N. Nagaosa,
\href{https://journals.aps.org/prb/abstract/10.1103/PhysRevB.98.054510}{Phys. Rev. B \textbf{98}, 054510 (2018)}.
{\bibitem{BookSD}
S. T. Ruggiero and D. A. Rudman (Eds.), \textit{Superconducting Devices} (Academic Press, Boston, 1990).
}
\bibitem{p38}
Q.-F. Sun, J. Wang, and H. Guo,
\href{https://journals.aps.org/prb/abstract/10.1103/PhysRevB.71.165310}{Phys. Rev. B \textbf{71}, 165310 (2005)}.
\bibitem{p39}
Y. Zhu, Q.-F. Sun, and T.-H. Lin,
\href{https://journals.aps.org/prb/abstract/10.1103/PhysRevB.66.085306}{Phys. Rev. B \textbf{66}, 085306 (2002)}.
\bibitem{p42}
S.-G. Cheng and Q.-F. Sun,
\href{https://iopscience.iop.org/article/10.1088/0953-8984/20/50/505202/meta}{Journal of Physics: Condensed Matter \textbf{20}, 505202 (2008)}.
\bibitem{p40}
Q.-F. Sun, J. Wang, and T.-H. Lin,
\href{https://journals.aps.org/prb/abstract/10.1103/PhysRevB.59.3831}{Phys. Rev. B \textbf{59}, 3831 (1999)}.
\bibitem{p41}
Q.-F. Sun, J. Wang, and T.-H. Lin,
\href{https://journals.aps.org/prb/abstract/10.1103/PhysRevB.59.13126}{Phys. Rev. B \textbf{59}, 13126 (1999)}.
\bibitem{p43}
C. W. J. Beenakker,
\href{https://journals.aps.org/prl/abstract/10.1103/PhysRevLett.67.3836}{Phys. Rev. Lett. \textbf{67}, 3836 (1991)}.
\bibitem{p44}
P. F. Bagwell,
\href{https://journals.aps.org/prb/abstract/10.1103/PhysRevB.46.12573}{Phys. Rev. B \textbf{46}, 12573 (1992)}.
\bibitem{p45}
A. Krichevsky, M. Schechter, Y. Imry, and Y. Levinson,
\href{https://journals.aps.org/prb/abstract/10.1103/PhysRevB.61.3723}{Phys. Rev. B \textbf{61}, 3723 (2000)}.
\bibitem{Zhang2013_CPR}
S.-F. Zhang, W. Zhu, and Q.-F. Sun, \href{http://dx.doi.org/10.1088/0953-8984/25/29/295301}{J. Phys.: Condens. Matter \textbf{25}, 295301 (2013).}
{\bibitem{BookSSP}
N. W. Ashcroft and N. D. Mermin, \textit{Solid State Physics} (World Publishing Corp., 2004).
}
\bibitem{p47}
When phase $\phi$ does not have an impact on the other parameters in the Hamiltonian, the magnitude of discrete supercurrent is proportional to the derivative of the Andreev bound state with respect to $\phi$. However, since the phase $\phi$ has an impact on the current, it will affect the Hamiltonian $H(I)$, and the proportional relation is a bit deviated.
\bibitem{p48}
Q. Yan, Y.-F. Zhou, and Q.-F. Sun,
\href{https://iopscience.iop.org/article/10.1088/1674-1056/aba272/meta}{Chin. Phys. B \textbf{29}, 097401 (2020)}.
\bibitem{p49}
Q. Cheng, Q. Yan, and Q.-F. Sun,
\href{https://journals.aps.org/prb/abstract/10.1103/PhysRevB.104.134514}{Phys. Rev. B \textbf{104}, 134514 (2021)}.
\bibitem{p50}
Y. Tanaka, T. Yokoyama, and N. Nagaosa,
\href{https://journals.aps.org/prl/abstract/10.1103/PhysRevLett.103.107002}{Phys. Rev. Lett. \textbf{103}, 107002 (2009)}.
\bibitem{p51}
D. B. Szombati, S. Nadj-Perge, D. Car, S. R. Plissard, E. P. A. M. Bakkers, and L. P. Kouwenhoven,
\href{https://www.nature.com/articles/nphys3742}{Nature Physics \textbf{12}, 568 (2016)}.
\bibitem{p52}
M. Alidoust, M. Willatzen, and A.-P. Jauho,
\href{https://journals.aps.org/prb/abstract/10.1103/PhysRevB.98.085414}{Phys. Rev. B \textbf{98}, 085414 (2018)}.
\bibitem{p53}
M. Alidoust,
\href{https://journals.aps.org/prb/abstract/10.1103/PhysRevB.101.155123}{Phys. Rev. B \textbf{101}, 155123 (2020)}.
\bibitem{p54}
C. H. Back, C. Würsch, D. Kerkmann, and D. Pescia,
\href{https://link.springer.com/article/10.1007/BF01313008}{Zeitschrift für Physik B Condensed Matter \textbf{96}, 1 (1994)}.
\bibitem{p55}
P. K. Das, A. Bhattacharyya, R. Kulkarni, S. K. Dhar, and A. Thamizhavel,
\href{https://journals.aps.org/prb/abstract/10.1103/PhysRevB.89.134418}{Phys. Rev. B \textbf{89}, 134418 (2014)}.
\bibitem{Ding2021_chl}
S. Ding, Z. Liang, J. Yang, C. Yun, P. Zhang, Z. Li, M. Xue, Z. Liu, G. Tian, F. Liu, W. Wang, W. Yang, and J. Yang, \href{http://link.aps.org/doi/10.1103/PhysRevB.103.094429}{Phys. Rev. B \textbf{103}, 094429 (2021).}
\bibitem{p56}
J. F. Schenck,
\href{https://aapm.onlinelibrary.wiley.com/doi/abs/10.1118/1.597854}{Medical Physics \textbf{23}, 815 (1996)}.
{\bibitem{BookITS}
M. Tinkham, \textit{Introduction to Superconductivity}, 2nd ed. (McGraw-Hill, New York, 1996).
}
\end{thebibliography}
\end{document}